\documentclass[11pt]{article}
\usepackage{mathtools}
\usepackage{epsf,graphics}

\def\beq{\begin{equation}}
\def\eeq{\end{equation}}
\date{\today}

\begin{document}
\begin{center}
{\large\bf Fermionic duality }\\[.3in]
  {\bf   Nicolas Sourlas} \\
  Laboratoire de Physique de l'Ecole Normale Sup\'erieure, ENS, Universit\'e PSL, CNRS, Sorbonne Universit\'e, Universit\'e Paris-Diderot, Sorbonne Paris Cit\'e, Paris, France
   
\ 

\end{center}
\vskip .15in
\centerline{\bf ABSTRACT}
Duality transformations play a very important role in theoretical physics.
In this paper I propose new duality transformations for fermionic theories. 
They map the strong coupling regime of one theory to the weak coupling regime 
of another theory. These transformations are based on the functional integral representation 
of the fermionic theories in terms of Grassmann variables and rely heavily on 
the properties of Grassmann variables. Potential applications include the 
study of the strong coupling phase of the two dimensional Hubbard model.

\begin{quotation}


\end{quotation}
   
Since first invented by Dirac for electromagnetism\cite{D} and  
Kramers and  Wannier\cite{KW}  for statistical 
mechanics, dualities  played a very important role in theoretical physics. 
Usually duality transformations establish a correspondance between the strong coupling 
regime of one theory and the weak coupling regime of the same (selfduality) or a different theory. 
In recent years new  duality transformations has been proposed, 
particularly for string theories.

In the present paper I propose a new class of dualities. I establish a correspondance 
between the weak (strong) coupling regime of a theory of fermions with the strong (weak) 
coupling regime of another theory of fermions. The derivation of this relation is based on the 
functional integral representation of the theories and the properties of the Grassmann variables 
entering their functional integral representation.

It is well known that partition functions and correlation functions of 
fermionic theories can be written as functional integrals over 
anticommuting Grassmann variables\cite{F,S}.

Consider the case of two component fermions and their effective fermionic Lagrangian  
written in terms of the two component Grassmann variables $ \psi $ and $\overline{\psi} $.
\begin{equation}
\label{L}
\mathcal{L} =  \sum_{i,j} {\overline{\psi}}_{i} M_{ij} \psi_{j}  
+ \sum_{i} ({\overline{\psi}}_{i}  \psi_{i} )^2 \lambda / 2  
+ {\overline{\psi}}_{i} J_{i} + {\overline{J}}_{i} \psi_{i} 
\end{equation}
$i$ and $j$ are the sites of a regular lattice, and $ J_{i} $ and $\overline{J}_{i}$ 
are two component fermionic "external sources". 
$M_{ij}$ are two by two matrices defined on the links of the lattice. 
$M_{ij}$ includes the kinetic energy and the mass term.
The partition function is given as usually by
\begin{equation}
\label{Z}
Z=  \int_{ \psi,\overline{\psi} } \exp{(\sum_{i,j} {\overline{\psi}}_{i} M_{ij} \psi_{j}  
+ \sum_{i} ({\overline{\psi}}_{i}  \psi_{i} )^2 \lambda / 2  
+ {\overline{\psi}}_{i} J_{i} + {\overline{J}}_{i} \psi_{i})}
\end{equation}
The purpose of the external sources is the possibility of the computation of correlation functions 
as functional derivatives. As an example,

\begin{equation}
\label{cor}
 <\overline{\psi}(x) \psi(y)> = \partial_{J_x} \partial_{\overline{J}_y} \log Z |_{J = \overline{J} =0}  
\end{equation}

At each point of the lattice we introduce new two component Grassmann variables 
 $ \chi_{i} $ and $\overline{\chi}_{i}$.

Using the well known properties of integrals over Grassmann variables 
one can easily derive  the following identity,
written in terms of the two component Grassmann variables $ \psi $ and $\overline{\psi} $.
\begin{equation}
\label{i0}
 \int_{ \chi,\overline{\chi} }  \exp ( \frac{(\overline{\chi} \chi )^2  u}{2} + 
 m {\overline{\chi}} \chi + {\overline{\chi}} \psi + {\overline{\psi}} \chi ) =  
 (u+m^2)  \exp ( -\frac{m}{m+u^2} \overline{\psi} \psi + {\frac{u}{2 (m+u^2)}}
(\overline{\psi} \psi)^2  ) 
\end{equation} 
On can verify this identity by expanding the exponential on both sides of the 
equation, using the fact that all higher powers of the Grassmann variables 
are identically zero and that the only non vanishing integral is 
$ \int_{ \chi,\overline{\chi} } (\overline{\chi} \chi )^2 = 1  $.

We introduce new Grassmann variables $ \chi_i $ and $\overline{\chi}_{i} $ at 
each point of the lattice and 
 use the previous identity to replace  $ \exp({\overline{\psi}}_{i} \psi_{i} )^2 \lambda / 2 ) $
terms in Equation ~\eqref{i0} as integrals over $ \chi_i $ and ${\overline{\chi}}_{i} $.
\begin{equation}
\label{i00}
\exp { \frac{\lambda ( {\overline{\psi}}_{i}  \psi_{i} )^2   }{2} }= 
\int_{ \chi,\overline{\chi} }  \exp (  (\overline{\chi}_{i}  \chi_{i}  )^2  u/2 + 
 m {\overline{\chi}_{i} } \chi_{i}  + {\overline{\chi}_{i} } \psi_{i}  + {\overline{\psi}}_{i}  \chi_{i}   ) 
\end{equation}
 where $ \lambda = u / (u+m^2)^2 $.   We obtain
 \begin{equation}
\label{i1}
 \begin{multlined}
Z = \int_{ \psi,\overline{\psi} }  \exp ( \sum_{ij} \overline{\psi}_{i} M{ij} \psi_{j} + 
\sum_{i} (\overline{\psi}_{i}  \psi_{i} )^2 \lambda /2 
+ \overline{\psi}_{i} J _{i}+ \overline{J}_{i} \psi _{i}) =   {1 / (u+m^2)^N } \\  
 \int_{ \psi,\overline{\psi} ,\chi,\overline{\chi} } \exp ( \sum_{ij} \overline{\psi}_{i} M^{'}_{ij} \psi_{i} + 
\sum_{i} (\overline{\psi}_{i} J_{i} + \overline{J}_{i} \psi_{i}  + (\overline{\chi}_{i} \chi_{i})^2  u/2 + 
m \overline{\chi}_{i} \chi_{i} + \overline{\chi}_{i} \psi_{i} + \overline{\psi}_{i} \chi_{i} )) 
\end{multlined}
\end{equation}
$N$ is the number of lattice sites, $M^{'}$ a new matrix $  M^{'}_{ij}=
M_{ij} + \delta_{ij} m/(u+m^2)  $ and $ \lambda = u / (u+m^2)^2 $.

The integrals over the $\psi$ fields are Gaussian and can be explicitly carried out and we find
that the partition function 
\begin{equation}
\label{i2}
 Z= \int_{ \psi,\overline{\psi} }  \exp {\cal{L}}_{\psi} = {\frac{1 }{(u+m^2)^N }}
\det(M{'}) \exp( -\overline{J}(M^{'})^{-1} J )
\int_{ \chi,\overline{\chi} } \exp  {\cal{L}}_{\chi} 
\end{equation}
where  
\begin{equation}
\label{i3}
 {\cal{L}}_{\psi} = \overline{\psi}_{i} M_{ij} \psi_{j} + 
(\overline{\psi}_{i}  \psi_{i} )^2 \lambda /2 + \overline{\psi}_{i} J_{i} + \overline{J}_{i} \psi_{i} 
\end{equation}
\begin{equation}
\label{i4}
 {\cal{L}}_{\chi} = - \overline{\chi}_{i} ((M^{'})^{-1}_{ij}+m\delta_{ij})  \chi_{j} 
 + (\overline{\chi}_{i} \chi_{i})^2  u/2
-\overline{\chi}_{i} (M^{'})^{-1}_{ij}  J_{i} - \overline{J}_{i} (M^{'})^{-1}_{ij}   \chi_{j} 
\end{equation}
$N$ is the number of sites of the lattice. 

This is a duality transformation. In fact this is a family of transformations depending on the free 
parameter $m$, which can be either positive or negative. 
$ {\cal{L}}_{\psi} $ is the original action, 
while $ {\cal{L}}_{\chi} $ is the action of the dual theory.
An effective action with a quartic coupling 
$\lambda$ is transformed into another one with quartic coupling $u $. 
The relation between the  two couplings is 
\begin{equation}
\label{i5}
 \lambda = \frac{u } {(u+m^2)^2 }
\end{equation}

As usual for duality transformations it relates weak coupling to strong coupling regimes.
The relation between the quartic couplings $\lambda $ and $ u $ depends on the parameter $ m$.
For $m=0$ $ \lambda = 1 / u $. For fixed and finite $m$ the solution which maximizes 
$\lambda $ is  $u=m^2 $, $ \lambda = 1 / 4m^2 $. 
 Large $\lambda $ and small $u$ require small $m$.

This duality transformation maps a functional integral to another one. 
It is not a transformation of a Hamiltonian to a new Hamiltonian.
Any Hamiltonian system can be put in the form of a functional integral\cite{F}.
The general conditions  necessary for a fermionic functional integral to correspond to a 
Hamiltonian system are not known. In certain cases reflection positivity\cite{OS1,OS2} 
is sufficient.

Correlation functions are obtained, as usually, by  the derivatives of $\log Z $ 
with respect of the external sources $ J_{i} $ and $\overline{J}_{i}$ at $ J = \overline{J} =0 $. 
For example 
\begin{equation}
\label{i15}
 <\overline{\psi}_x \psi_y>_{\psi} = \partial J_x \partial \overline{J}_y \log Z |_{J = \overline{J} =0}  = 
 -(M^{'})^{-1}_{xy} +  <( M^{'})^{-1}_{xj}  \overline{\chi}_{j}  (M^{'})^{-1}_{iy} \chi_{i} > _{\chi} 
\end{equation} 
$<\cdots>_{\psi} $ is computed in the theory described by ${\cal{L}}_{\psi} $ while 
$<\cdots>_{\chi} $ is computed in the theory described by $ {\cal{L}}_{\chi}$.
The correlation functions of the original theory are thus related by this simple expression to the correlation 
functions of the dual theory.

Usually one is interested in Hamiltonians $H_{\psi}$ with short range interactions, i.e. 
the matrix $M$ is short range. What is the range of interactions of the dual theory, i.e. 
what is the range of the matrix $ M^{''}_{ij} = (M^{'})^{-1}_{ij}+m\delta_{ij} $ ? 

We assume translation invariance, i.e.  $M_{xy} = g(x-y) $,
 in which case the matrices $M$ and $M^{'} = M + I m/(u+m^2) $ 
 ($ I $ is the unit matrix), can be diagonalized by Fourier transform. 
If $M_{xy} = \int_k \exp(-ik(x-y)) f(k) $, 
\begin{equation}
\label{i6}
((M^{'})^{-1})_{xy} = \int_k {\frac{\exp( ik(x-y) ) } { f(k)+ m/(m+u^2)} }
\end{equation} 
It is well known that the behaviour of $M_{xy} $ for large separations of $ x-y $ is 
governed by the singularities in $k$ space in the previous equation for small $k$, i.e. the zeros 
of the denominator.
If we have cubic symmetry, the first terms of the expansion of $ f(k) $ in powers of $k$ 
$ f(k) =f_0 +f_1 \vec{k}^2 + \cdots $ 
are rotation invariant. The closest singularity to the origin in Equation ~\eqref{i6} is 
for $ k^2 = -\mu $, $\mu= (f_0 + m/(m+u^2))/f1 $ and $ M^{''}_{xy} \sim \exp(-|x-y| \mu )$ 

Because $m$ is a free parameter of the transformation, which can be either positive or negative, 
the interaction range of $ M^{''} $ $ 1/ \mu $ is adjustable. 
We can even choose $\mu \sim 0$, i.e. $ m \sim - u f_0 / (f_0 +1) $ in which the dual action ${\cal{L}}_{\chi} $
is infinite range! 
In that case the behavior of the theory described by ${\cal{L}}_{\chi} $ can be computed in 
the mean field theory approximation.

Because of equation ~\eqref{i5} the choice of the value of $ m $ affects also the relation between 
the couplings $\lambda $ and $u$.
We would be interested to study the large $\lambda $ coupling regime of the original 
theory by mapping it to a week $ u $ coupling dual theory with short range interactions, i.e. 
large value of $ \mu $ and small value of $ u $. Is this possible?

Suppose the original theory has nearest neighbor couplings on a cubic lattice in $D$ dimensions. 
$$ M_{x,y} = r \delta (\vec{x},\vec{y}) + J ( \delta(\vec{x}, \vec{y}+\vec{\mu}) + \delta(\vec{x}, 
\vec{y}-\vec{\mu}) )     $$
where $ \delta(x,y)$ is the Kronecker $\delta $ function and $ \vec{\mu}  $ the unit vector 
along the axis's of the lattice.
The Fourier transform is 
$$ M_F (k) = r + J \sum_{\vec{\mu}} \cos(k_{\mu}) = r+DJ + J  (\vec{k}^2 ) / 2 + 
 J/24  \sum_{\vec{\mu}} k_{\mu}^4 + \cdots $$ 
\begin{equation}
\label{i8}
((M^{'})^{-1})_{xy} = \int_k {\frac{\exp( i\vec{k}(\vec{x}-\vec{y}) ) } {r+DJ + m/(m+u^2) + J  \vec{k}^2  /2
} }  \sim \exp(- \mu |\vec{x}-\vec{y} | )
\end{equation} 
where $ \mu = 2 (r+D J + m/(m+u^2)) / J$.  We showed above that for finite value of the 
parameter $ m $, $ \lambda  \le 1/4m^2 $. If $ \lambda  = 1/4m^2 $, 
  $u=m^2$ and 
$ \mu = 2(r+D J + 2/(1+m^3)  ) $. 
We conclude that for small $m$ the original large coupling theory 
is mapped to a weak coupling theory with short range interactions.

These duality transformation are valid for any dimension of space, 
provided the fermions have only two components. If $\chi $ and  $\psi $ 
have more than two components equation ~\eqref{i0} becomes more complicated:
higher powers of $\overline{\psi} \psi  $ appear. If we are close to a Gaussian fixed point 
of the renormalization group, we may neglect these terms 
as being irrelevant in the renormalization group sense.

Of particular interest in condensed matter physics is the Hubbard model\cite{H}.
 It is used to describe the properties of strongly correlated electron systems.
The Hamiltonian of the Hubbard model is
\begin{equation}
\label{i10}
H = {\sum_{\vec{r} {\vec{r}}^{'} } } -t (b^{+}_{\vec{r}} b_{\vec{r}^{'}} + 
{b^{+}}_{\vec{r}^{'}} b_{\vec{r}} ) + 
\mu \sum_{\vec{r}} b_{\vec{r}}^{+} b_{\vec{r}}
+ {u / 2}  \sum_{\vec{r}} (b_{\vec{r}}^{+} b_{\vec{r}})^2   
\end{equation}
$b_{\vec{r}}^{+} $ and   $ b_{\vec{r}} $ are the creation and annihilation 
operators of spin $1/2$ electrons.

There is a standard procedure\cite{F,SCH,S} to write the statistical mechanics partition function 
 of fermionic systems $ Z= Tr \exp (-\beta H ) $ 
as a functional integral over Grassmann variables by using the identity 
$ Z= Tr \exp (-\beta H ) = Tr ((\exp (-\epsilon H ) )^K ) $, 
where $ \epsilon = \beta / K $ and taking the large $ K$ limit.
One introduces an additional imaginary time coordinate $i = 1,\cdots, K$ and the 
 Grassmann variables are functions of $\vec{x}$ and $i$.
 
 The partition function as afunctional integral is then 
 \begin{equation}
\label{i11}
 Z= \int_{ \psi,\overline{\psi} } \exp (\sum_{1}^{K} \sum_{\vec{r} }
 (\overline{\psi}_{i}(\vec{r} ) -\overline{\psi}_{i-1}(\vec{r} )) 
\psi_{i}(\vec{r} ) )  -\epsilon  H (\overline{\psi}_{i} , \psi_{i} ) )
\end{equation}
where 
\begin{equation}
\label{i12}
 H = {\sum_{\vec{r} {\vec{r}}^{'} }}  -t (\overline{\psi}_{\vec{r}} \psi_{\vec{r}^{'}} + 
{\overline{\psi}}_{\vec{r}^{'}} \psi_{\vec{r}} ) + 
\mu \sum_{\vec{r}} \overline{\psi}_{\vec{r}} \psi_{\vec{r}}
+ {u / 2}  \sum_{\vec{r}} (\overline{\psi}_{\vec{r}} \psi_{\vec{r}})^2 
\end{equation}
The only difference with the previous case is in the form of the matrix $M$.
\begin{equation}
\label{i13}
\overline{\psi}_{i} ({\vec{r}} ) M(i,\vec{r}; i^{'},\vec{r}^{'} ) \psi_{{i}^{'}} ({\vec{r}}^{'} )
=(\overline{\psi}_{i} ({\vec{r}}) 
 -\overline{\psi}_{i-1} ({\vec{r}} ) ) \psi_{i} ({\vec{r}} ))  -
t \epsilon (\overline{\psi}_{i}({\vec{r}}) \psi_{i}({\vec{r}^{'}}) + 
{\overline{\psi}}_{i}({\vec{r}^{'}}) \psi_{i}({\vec{r}} ) ) + 
\mu  \epsilon \overline{\psi}_{i}({\vec{r}}) \psi_{i}({\vec{r}})
\end{equation}
or in Fourier space
\begin{equation}
\label{i14}
 M(\vec{k},\omega) = 1 - \exp (-i \omega )  + 2 t \epsilon \sum_{\mu} cos (k_{\mu} ) + 
 \mu \epsilon
\end{equation}

 The phase structure of the Hubbard model in the strong coupling regime is 
still debated. It would be very interesting to use this new duality transformation, outlined above,  
to transform the strong coupling Hubbard model to a weak coupling model and with the help of 
perturbation theory  clarify this important issue.

A different dual approach\cite{G} has already been proposed for the two-dimensional Hubbard model.

\end{document}